\documentclass[aps,prb,twocolumn,nobibnotes]{revtex4-2}  
\usepackage{dcolumn}
\usepackage{graphicx}
\usepackage{color}
\usepackage{amssymb}
\usepackage{amsmath}
\usepackage{bm}
\usepackage[colorlinks=true,linktocpage=true,breaklinks=true]{hyperref}
\usepackage{multirow}
\usepackage{braket}
\usepackage{appendix}

\begin{document}

\title{Charge pumping with strong spin-orbit coupling:\\ Fermi surface breathing, Berry curvature and higher harmonic generation}

\author{Aur\'{e}lien Manchon}
\email{email: aurelien.manchon@univ-amu.fr}
\author{Armando Pezo}
\affiliation{Aix-Marseille Universit\'e, CNRS, CINaM, Marseille, France}

\date{\today}

\begin{abstract}
Spin and charge pumping induced by a precessing magnetization has been instrumental to the development of spintronics. Nonetheless, most theoretical studies so far treat the spin-orbit coupling as a perturbation, which disregards the competition between exchange and spin-orbit fields. In this work, based on Keldysh formalism and Wigner expansion, we develop an adiabatic theory of spin and charge pumping adapted to systems with arbitrary spin-orbit coupling. We apply this theory to the magnetic Rashba gas and magnetic graphene cases and discuss the pumped ac and dc current. We show that the pumped current possesses both intrinsic (Berry curvature-driven) and extrinsic (Fermi surface breathing-driven) contributions, akin to magnetic damping. In addition, we find that higher harmonics can be generated under large-angle precession and we propose a couple of experimental setups where such an effect can be experimentally observed. 
\end{abstract}

\maketitle

\section{Introduction} 
Adiabatic spin pumping \cite{Brataas2002,Tserkovnyak2002b} has been instrumental to the development of spintronics over the past two decades. It is now routinely used to inject pure spin currents from a magnetic spin source into an adjacent metal, enabling the investigation of spin-to-charge interconversion processes in a wide range of materials, from transition metal compounds \cite{Saitoh2006} to two-dimensional gases \cite{Rojas-Sanchez2013b}, oxide heterostructures \cite{Vaz2019}, topological surface states \cite{Mellnik2014,Jamali2015,Rojas-Sanchez2016b}, van der Waals heterostructures (see for instance Ref. \cite{Galceran2021}) but also other magnetic materials such as antiferromagnets \cite{Frangou2016} and spin liquids \cite{Hirobe2017}. Although the magnetic spin source is usually a ferromagnet excited at magnetic resonance, the recent demonstration of spin-to-charge interconversion using antiferromagnetic resonance \cite{Li2020c,Vaidya2020} opens appealing avenues for the generation of very high-frequency currents via spin pumping. In the standard theory of spin pumping \cite{Brataas2002,Tserkovnyak2002b}, the interfacial spin current induced by the precessing magnetization ${\bf m}$ and injected in the adjacent metal reads
\begin{eqnarray}\label{eq1}
{\cal J}_s=\eta_r {\bf m}\times\partial_t {\bf m}+\eta_i\partial_t {\bf m},
\end{eqnarray}
where $\eta_{r,i}$ are coefficients related to the spin mixing conductance at the interface between the magnet and the nonmagnetic metal, and to the spin relaxation in the metal \cite{Brataas2001,Tserkovnyak2003}. The polarization of the spin current is along the magnetization vectors on the right-hand side of Eq. \eqref{eq1}. When spin-orbit coupling is present in the metal, the spin current is converted into a charge current that takes the general form
\begin{eqnarray}\label{eq2}
{\bf J}_c=\alpha_H\eta_r {\bf z}\times({\bf m}\times\partial_t {\bf m})+\alpha_H\eta_i{\bf z}\times\partial_t {\bf m},
\end{eqnarray}
where ${\bf z}$ is normal to the interface and $\alpha_H$ is the spin-to-charge conversion efficiency, proportional to the spin-orbit coupling strength, and whose specific structure depends on the involved mechanism (spin Hall effect \cite{Dyakonov1971b,Sinova2015}, Rashba-Edelstein effect \cite{Edelstein1990,Manchon2015}, possibly spin swapping \cite{Lifshits2009} etc.). Equation \eqref{eq2} is widely used to interpret experimental data and quantify the physical parameters such as the spin-mixing conductance itself, the spin-to-charge conversion efficiency and the spin relaxation length \cite{Saitoh2006,Saitoh2012a,Mosendz2010,Weiler2013,Rojas-Sanchez2014}. Notice that, to date, the vast majority of experiments have focused on the time-averaged, rectified part of the pumped charge current ${\bf J}_c|_{\rm dc}=\alpha_H\eta_r \langle{\bf z}\times({\bf m}\times\partial_t {\bf m})\rangle$, and only a handful of them have achieved to measure the ac contribution \cite{Hahn2013,Weiler2014,Wei2014}.

\begin{figure}[h!]
  \includegraphics[width=9cm]{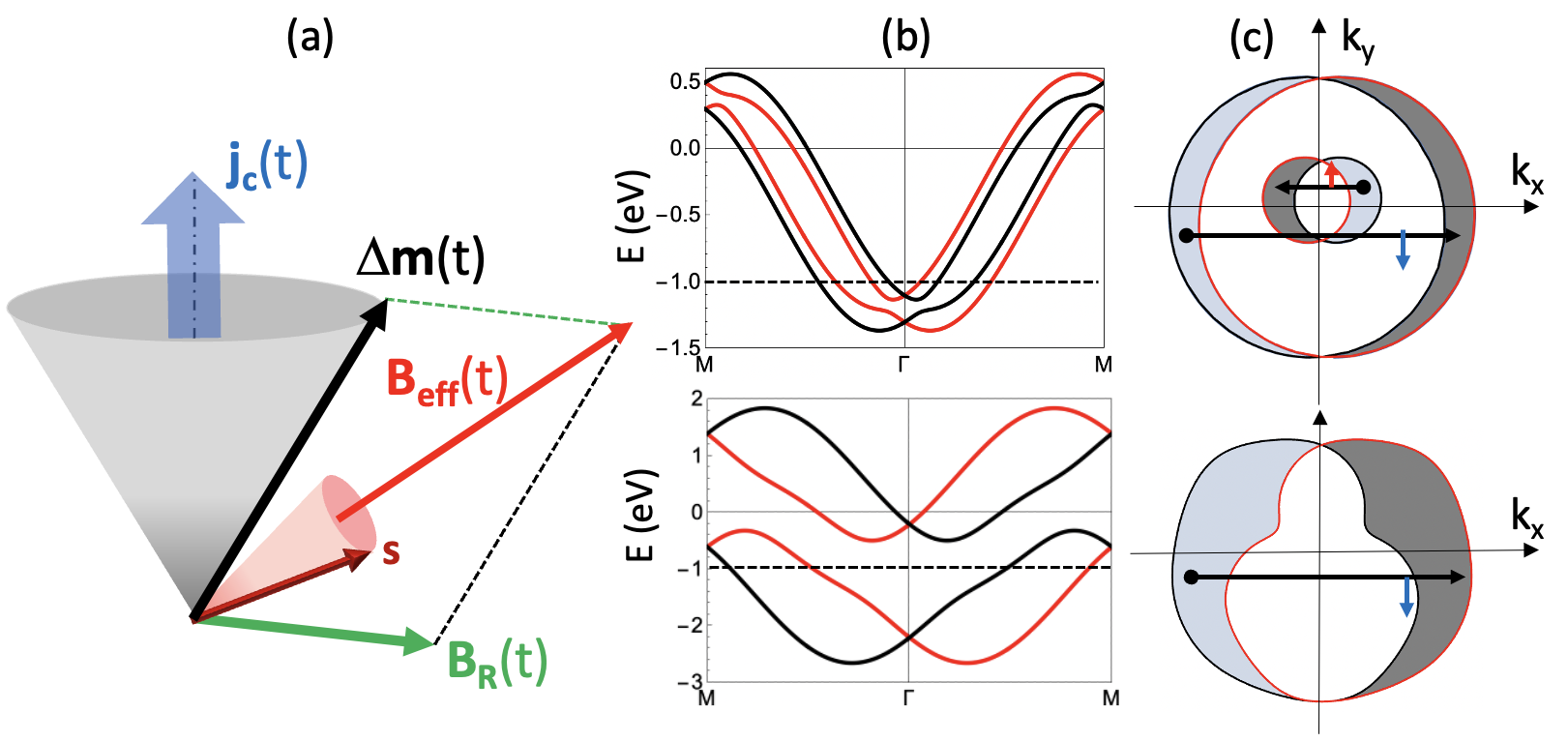}
  \caption{(Color online) (a) Sketch of the nonlinear spin dynamics expected when a strong Rashba spin-orbit coupling coexists with s-d exchange. (b) Band structure of the magnetic Rashba gas for ${\bf m}={\bf y}$ (red) and $-{\bf y}$ (black) for (top) $\Delta=0.1t$, $t_R=0.1t$ and (bottom) $\Delta=1t$, $t_R=0.3t$. (c) Corresponding Fermi surface computed at $E_F=-1t$, showing a strong "breathing" effect upon magnetization reorientation. In these calculations, $t=0.2$. \label{fig:sketch}}
\end{figure}

An important shortcoming of the standard theory of spin pumping based on the spin mixing conductance is that it formally applies in the presence of vanishingly small spin-orbit coupling compared to the s-d exchange between conduction and localized electrons (about 1-2 eV in Fe, Co, Ni compounds). In fact, in most experiments, the adjacent metal rather possesses a large spin-orbit coupling, i.e., a few 100 meV (heavy metals, topological insulators, and Weyl semimetals, to name a few). In other words, the spin mixing conductance approach is not adapted to treat these systems and overlooks the competition between exchange and spin-orbit interactions. As a matter of fact, in noncentrosymmetric multilayers interfacial spin-orbit splitting adopts the form of Rashba spin-orbit coupling \cite{Manchon2015}, which substantially modifies the spin dynamics at the interface. As sketched in Fig. \ref{fig:sketch}(a), the time-dependent current ${\bf j}_c(t)$ pumped by the precessing magnetization is accompanied by the so-called Rashba field \cite{Edelstein1990,Manchon2015}, ${\bf B}_R(t)\propto {\bf z}\times{\bf j}_c(t)$, which competes with the s-d exchange to drag the itinerant spin away from the magnetization. Since this Rashba field is itself proportional to the charge current, one naturally expects a massive modification of the spin dynamics in the limit of strong spin-orbit coupling. 

To properly model the competition between exchange interaction and spin-orbit coupling, both effects must be treated on equal footing. This was, for instance, achieved by Mahfouzi et al. \cite{Mahfouzi2010,Mahfouzi2012} using the nonequilibrium Green's function approach (see also Ref. \cite{Dolui2022}). In Ref. \cite{Chen2015c}, Chen and Zhang proposed a Green's function approach to compute the induced spin current in the presence of spin-orbit coupling, although limited to small precession angles and focusing on the dc charge current. More recently, in Ref. \cite{Ly2022}, we performed time-dependent quantum transport simulations and reported the progressive emergence of higher harmonics upon increasing the strength of Rashba spin-orbit coupling. This result, confirmed by Ref. \cite{VarelaManjarres2023}, clearly advocates for the presence of nonlinear itinerant spin dynamics but its lack of transparency hinders a precise understanding of the underlying physics. 

In the present work, we address this problem by adopting a different theoretical approach. Based on Keldysh formalism, we first derive a formula for the charge pumping that is valid in the slow dynamics regime and, most importantly, valid for the full range of spin-orbit coupling and exchange interaction. We then apply this formalism to the magnetic Rashba gas and magnetic graphene, and demonstrate that the harmonic generation is a direct consequence of the Berry curvature in mixed time-momentum space. We then suggest several materials systems and configurations in which the harmonic generation can be observed.

\section{Adiabatic pumping, Fermi surface breathing, and Berry curvature}
\subsection{Keldysh theory of adiabatic pumping} 

Let us start from Keldysh-Dyson equation \cite{Rammer1986} in Wigner representation, i.e., only the macroscopic coordinates (time and position) of the center-of-mass of the wave packet are treated explicitly while its microscopic internal degrees of freedom are Fourier transformed (see, e.g., Ref. \cite{Hajr2020}). In the present theory, we are interested in deriving the response of an observable ${\cal O}$, expressed through the lesser Green's function $G^<_{\bf k}$, to the first order in magnetization dynamics, $\partial_t{\cal H}_{\bf k}$. Keldysh-Dyson equations can be rewritten
\begin{eqnarray}\label{eq:keldysh1}
(\varepsilon-{\cal H}_{\bf k}-\Sigma^R)\otimes G^R_{\bf k}=1,\\
G^<_{\bf k}=G^R_{\bf k}\otimes\Sigma^<\otimes G^A_{\bf k},
\end{eqnarray}
where ${\cal H}_{\bf k}$ is the unperturbed Hamiltonian, $\otimes={\rm exp}[i\hbar(\overleftarrow{\partial}_t\overrightarrow{\partial}_\varepsilon-\overleftarrow{\partial}_\varepsilon\overrightarrow{\partial}_t)]$ is Moyal's product that emerges from Wigner transform, and $\Sigma^{R,<}=n_iV_0^2\int d^3{\bf k}/(2\pi)^3G_{\bf k}^{R,<}$ is the (retarded, lesser) self-energy in the presence of delta-like impurities with potential $V_0$ and density $n_i$. We now expand these two equations to the first order in $\partial_t{\cal H}_{\bf k}$ and after some algebra, one obtains
\begin{eqnarray}
G^<_{\bf k}&=&G^R_{\bf k}\Sigma^<G^A_{\bf k}-\frac{i\hbar}{2}\left(G^R_{{\bf k}0}\partial_t{\cal H}_{\bf k}\partial_\varepsilon G^A_{{\bf k}0}-\partial_\varepsilon G^R_{{\bf k}0}\partial_t{\cal H}_{{\bf k}}G^A_{{\bf k}0}\right),\nonumber\\\label{eq:keldysh2}\\
G^R_{\bf k}&=&G^R_{{\bf k}0}-\frac{i\hbar}{2}\left(G^R_{{\bf k}0}\partial_t{\cal H}_{{\bf k}}\partial_\varepsilon G^R_{{\bf k}0}-\partial_\varepsilon G^R_{{\bf k}0}\partial_t{\cal H}_{{\bf k}}G^R_{{\bf k}0}\right)\label{eq:gr}
\end{eqnarray}
We then simply insert Eq. \eqref{eq:gr} into Eq. \eqref{eq:keldysh2}, and compute the response of an observable ${\cal O}$ as
\begin{eqnarray}\label{eq:linear1}
{\cal O}=\int \frac{d\varepsilon}{2i\pi}{\rm Tr}_{\bf k}[\hat{O}G^<_{\bf k}].
\end{eqnarray}
Here, ${\rm Tr}_{\bf k}[...]=\int \frac{d^3{\bf k}}{(2\pi)^3}{\rm Tr}[...]$. By posing ${\cal O}=\delta{\cal O}_j\partial_tm_j$, and $\partial_t{\cal H}_{\bf k}=-{\bf T}_{\bf k}\cdot\partial_t{\bf m}$, where ${\bf T}_{\bf k}=-\partial_{\bf m}{\cal H}_{\bf k}$ is the torque operator, we obtain
\begin{eqnarray}\label{eq:resp1}
\delta{\cal O}_{j}&=&\langle\hat{O},T_{\bf k}^j\rangle,\\
&=&\hbar\int \frac{d\varepsilon}{2\pi}{\rm Re Tr}_{\bf k}[\hat{O}(G^R_{{\bf k}0}-G^A_{{\bf k}0})T_{\bf k}^jG^A_{{\bf k}0}]\partial_\varepsilon f(\varepsilon),\nonumber\\
&&+\hbar\int \frac{d\varepsilon}{2\pi}{\rm ReTr}_{\bf k}[\hat{O}G^A_{{\bf k}0}[T_{\bf k}^j,G^A_{{\bf k}0}]G^A_{{\bf k}0}]f(\varepsilon).
\end{eqnarray}
To separate explicitly the contribution that is even under time-reversal symmetry from the one that is odd, we adopt the symmetrization procedure discussed in Ref. \cite{Bonbien2020}, i.e., $\delta{\cal O}^{\rm surf}_{j}=(\langle\hat{O},T_{\bf k}^j\rangle+\langle T_{\bf k}^j,\hat{O}\rangle)/2$ and  $\delta{\cal O}^{\rm sea}_j=(\langle\hat{O},T_{\bf k}^j\rangle-\langle T_{\bf k}^j,\hat{O}\rangle)/2$. By definition, the former response is even under time reversal, whereas the latter is odd. They explicitly read

\begin{eqnarray}\label{eq:resp3}
\delta{\cal O}^{\rm surf}_{j}=-\hbar\int \frac{d\varepsilon}{4\pi}{\rm Re Tr}_{\bf k}[\hat{O}G^{R-A}_{{\bf k}0}T_{\bf k}^jG^{R-A}_{{\bf k}0}]\partial_\varepsilon f(\varepsilon),\\
\delta{\cal O}^{\rm sea}_{j}=\hbar\int \frac{d\varepsilon}{2\pi}{\rm ReTr}_{\bf k}[\hat{O}G^{R+A}_{{\bf k}0}T_{\bf k}^j\partial_\varepsilon G^{R-A}_{{\bf k}0}]f(\varepsilon),\label{eq:resp4}
\end{eqnarray}
with $G^{R\pm A}_{{\bf k}0}=G^{R}_{{\bf k}0}\pm G^{A}_{{\bf k}0}$. These two expressions represent the Fermi surface and Fermi sea contributions to the adiabatic pumping. The first term is associated with the time-dependent deformation of the Fermi surface, an effect called "Fermi surface breathing" \cite{Hodges1967} and that is proportional to extrinsic, impurity-driven momentum relaxation time (like, i.e., longitudinal conductivity). In the limit of weak disorder, the second term is associated with the Berry curvature in mixed time-momentum space [see Eq. \eqref{eq:bcformula}], as discussed further below, and is usually referred to as "intrinsic".

It is noteworthy to point out that this theory generalizes the theory of magnetic damping proposed by Gilmore et al. \cite{Gilmore2007}, in which intrinsic and extrinsic electronic contributions to magnetic damping are discussed. In our formalism, the magnetic damping can be computed simply by replacing $\hat{O}$ by the torque operator, resulting in the torque-torque correlation introduced by Kambersky \cite{Kambersky1976}. These formulas are valid for slow dynamics but, most importantly, remain exact for all values of spin-orbit coupling and exchange, as well as for any direction of the magnetization vector. It is also well adapted to multiband systems and heterostructures and can be used to compute spin, charge, and orbital pumping in realistic materials.


\subsection{Berry curvature and Fermi surface breathing} 

The charge current pumped by the precessing magnetization is obtained by replacing $\hat{O}$ by $-e\hat{\bf v}$, and we get
\begin{eqnarray}
{\bf J}_c^{\rm surf}&=&-e\hbar\int \frac{d\varepsilon}{4\pi}{\rm Re Tr}_{\bf k}[\hat{\bf v}G^{R-A}_{{\bf k}0}\partial_t{\cal H}_{\bf k}G^{R-A}_{{\bf k}0}]\partial_\varepsilon f(\varepsilon),\nonumber\\\label{eq:jcsurf}\\
{\bf J}_c^{\rm sea}&=&e\hbar\int \frac{d\varepsilon}{2\pi}{\rm ReTr}_{\bf k}[\hat{\bf v}G^{R+A}_{{\bf k}0}\partial_t{\cal H}_{\bf k}\partial_\varepsilon G^{R-A}_{{\bf k}0}]f(\varepsilon).\nonumber\\\label{eq:jcsea}
\end{eqnarray}
In the limit of slow dynamics, the electron's spin remains aligned on the effective field due to exchange and spin-orbit coupling [Fig. \ref{fig:sketch}(a)] and, because of spin-momentum locking, the wave function acquires a geometrical phase \cite{Sundaram1999,Xiao2010b},
\begin{eqnarray}\label{eq:bcformula}
{\bm\Omega}^n_{t{\bf k}}=2i{\rm Im}\left[\langle\partial_t n|\partial_{\bf k}n\rangle\right].
\end{eqnarray}
Following the semiclassical theory developed by Sundaram and Niu \cite{Sundaram1999}, this geometrical phase results in an intrinsic charge current. In the relaxation time approximation, by taking $G^{R,A}_{{\bf k}0}=\sum_n| n\rangle\langle n|/(\varepsilon-\varepsilon_{\bf k}^n\pm i\Gamma)$ with $\Gamma\rightarrow0$, it is straightforward to demonstrate that Eq. \eqref{eq:jcsea} reduces to,
\begin{eqnarray}
{\bf J}_c^{\rm int}=-e\sum_n\int \frac{d^2{\bf k}}{(2\pi)^2}{\bm\Omega}^n_{t{\bf k}}f(\varepsilon_{\bf k}^n).\label{bc-formula}
\end{eqnarray}
In addition to the Berry curvature, the high sensitivity of the Fermi surface to the magnetization direction results in a so-called "breathing", i.e., the periodic modulation of the Fermi surface driven by the precessing magnetization \cite{Hodges1967}. During the breathing, states of opposite spin chirality are pumped from one side of the Fermi surface to the other, resulting in a periodic charge current, see Fig. \ref{fig:sketch}(c). This effect is accounted for by Eq. \eqref{eq:jcsurf} and only involves intraband transitions. In the weak disorder limit, the charge current reduces to
\begin{eqnarray}
{\bf J}_c^{\rm ext}=-\frac{e\hbar}{2\Gamma}\sum_{nn'}\int \frac{d^2{\bf k}}{(2\pi)^2}\langle n|\hat{\bf v}| n\rangle\langle n|\partial_t{\cal H}_{\bf k}|n\rangle\delta(\varepsilon_F-\varepsilon_{\bf k}^n).\nonumber\\\label{fs-formula}
\end{eqnarray}
Notice that this Fermi surface breathing is also at the origin of electron-mediated magnetic damping \cite{Kunes2002,Fahnle2006,Gilmore2007}. As discussed in detail in the next section, whereas the Fermi surface remains mostly rigid when the spin-orbit coupling is small [top panels in Fig. \ref{fig:sketch}(c)],  its distortion upon magnetization precession becomes more pronounced for large spin-orbit coupling [bottom panels in Fig. \ref{fig:sketch}(c)], manifesting the drastic competition between spin-orbit coupling and exchange. 

\section{dc and ac charge pumping} 

In this section, we first discuss the features of charge pumping in the limit of weak spin-orbit coupling by comparing the current generated via the inverse spin Hall effect \cite{Dyakonov1971b,Sinova2015} and the Rashba-Edelstein effect \cite{Edelstein1990,Manchon2015}. We then move on to the general case of arbitrarily strong spin-orbit coupling by investigating the charge pumping in two standard situations, the paradigmatic magnetic Rashba gas and the magnetic graphene monolayer. The magnetic Rashba gas is the minimal model for magnetic systems with inversion symmetry-breaking \cite{Sinova2004,Manchon2008,Kim2013b} and, as such, is an excellent platform to establish salient features of charge pumping. On the other hand, graphene is a material of choice for charge pumping because of its long spin relaxation length \cite{Tombros2007} and its ability to acquire both magnetism and spin-orbit coupling by proximity effect \cite{Benitez2018,Safeer2019}. Several experiments have demonstrated spin pumping into heterostructures involving a graphene monolayer \cite{Mendes2015,Dushenko2016,Evelt2017,Leutenantsmeyer2017,Indolese2018}. The main difference between these two models is their energy dispersion, which is quadratic in the former and linear in the latter, resulting in distinct intrinsic and extrinsic current contributions.

\subsection{Preliminary insights}

Without loss of generality, we consider a magnetization precessing around the ${\bf x}$ axis (see Fig. \ref{fig:pump1}), ${\bf m}=(\cos\theta,\sin\theta\sin\phi,-\sin\theta\cos\phi)$, with $\phi=\omega t$. In the standard case of a ferromagnet/nonmagnetic metal bilayer, studied in most of the literature \cite{Saitoh2006,Saitoh2012a,Mosendz2010,Weiler2013,Rojas-Sanchez2014}, the charge pumping is due to the spin Hall effect in the nonmagnetic metal and computed by combining the magnetoelectronic circuit theory with the drift-diffusion equation \cite{Brataas2002,Tserkovnyak2002b}. In this case, the charge current reads
\begin{eqnarray}
{\bf J}_c\approx{\tilde \alpha}_N\frac{{\tilde \lambda}_N}{d_N}\left(2{\tilde G}^r_{\uparrow\downarrow}{\bf z}\times({\bf m}\times\partial_t{\bf m})+2{\tilde G}^i_{\uparrow\downarrow}{\bf z}\times\partial_t{\bf m}\right),
\end{eqnarray}
where 
\begin{eqnarray}
2{\tilde G}^{r,i}_{\uparrow\downarrow}&=&\frac{2G^{r,i}_{\uparrow\downarrow}}{1+\frac{2G_{\uparrow\downarrow}\tilde{\lambda}_N}{\sigma_N}},\; {\tilde \lambda}_N=\frac{\lambda_N}{\tanh\frac{d_N}{\lambda_N}},\\
{\tilde \alpha}_N&=&\alpha_N\left(1-\cosh^{-1}\frac{d_N}{\lambda_N}\right).
\end{eqnarray}
$G^{r,i}_{\uparrow\downarrow}$ are the real and imaginary parts of the spin mixing conductance, and $\alpha_N$, $\lambda_N$, $d_N$ and $\sigma_N$ are the spin Hall angle, spin relaxation length, thickness and conductivity of the nonmagnetic metal. For the precessing magnetization adopted in our calculations and considering ${G}^r_{\uparrow\downarrow}\gg{G}^i_{\uparrow\downarrow}$ \cite{Zwierzycki2005}, 
\begin{eqnarray}\label{eq:spt1}
{\bf J}_c&=&\omega{\tilde \alpha}_N\frac{{\tilde \lambda}_N}{d_N}2{\tilde G}^r_{\uparrow\downarrow}\left(\sin\theta\cos\theta\sin\omega t {\bf x}+\sin^2\theta {\bf y}\right)\\
&&+\omega{\tilde \alpha}_N\frac{{\tilde \lambda}_N}{d_N}2{\tilde G}^i_{\uparrow\downarrow}\sin\theta\cos\omega t{\bf x}.\nonumber
\end{eqnarray}
As mentioned in the introduction, this theory gives the well-known result that a dc current is injected transverse to the precession axis and an ac current is generated along it. Let us now turn our attention to the case of a magnetic Rashba gas.

Let us now compute this current in the case of the free magnetic Rashba electron gas, defined by the Hamiltonian
\begin{eqnarray}\label{Hrashba}
{\cal H}=\frac{\hbar^2k^2}{2m}+\Delta \hat{\bm\sigma}\cdot{\bf m}+\alpha_{\rm R} \hat{\bm\sigma}\cdot({\bf p}\times{\bf z}),
\end{eqnarray}
where $\alpha_{\rm R}$ is the Rashba strength. The eigenstates are 
\begin{eqnarray}\label{wfR}
|+\rangle=\left(\begin{matrix}-e^{-i\phi_k}\cos\frac{\theta_k}{2}\\\sin\frac{\theta_k}{2}\end{matrix}\right),\;|-\rangle=\left(\begin{matrix}e^{-i\phi_k}\sin\frac{\theta_k}{2}\\\cos\frac{\theta_k}{2}\end{matrix}\right),\\
\varepsilon_{\bf k}^n=\frac{\hbar^2k^2}{2m}+n\lambda_k,\; n=\pm1,
\end{eqnarray}
with
\begin{eqnarray}\label{wfR}
&&\lambda_k=\sqrt{\Delta^2+\alpha_{\rm R}^2k^2+2\Delta\alpha_{\rm R}({\bf p}\times{\bf z})\cdot{\bf m}}\\
&&\cos\theta_k=-\frac{\Delta}{\lambda_k}\sin\theta\cos\phi,\;\tan\phi_k=\frac{\Delta\sin\theta\sin\phi-\alpha_{\rm R}k_x}{\Delta\cos\theta+\alpha_{\rm R}k_y}.\nonumber\\
\end{eqnarray}

Although the scattering matrix formalism used in Refs. \cite{Brataas2002,Tserkovnyak2002b} is well adapted to current-perpendicular-to-plane geometries, it is not suited to current-in-plane geometries. To obtain an analytical expression of the pumped charge current in the two-dimensional gas, we rather use Eqs. \eqref{bc-formula} and \eqref{fs-formula}. In the Rashba gas, the Berry curvature for band $n$ reads
\begin{eqnarray}\label{bctk}
{\bm \Omega}_{t{\bf k}}^n=n(\partial_{\bf k}\theta_k\partial_t\phi_k-\partial_{\bf k}\phi_k\partial_t\theta_k)\frac{\sin\theta_k}{2}.
\end{eqnarray}
After some algebra, we find that the total pumped current density reads, to the lowest order in Rashba strength $\alpha_{\rm R}$,
\begin{eqnarray}\label{pumpedRash}
{\bf J}_c&=&-\frac{e\omega}{\lambda_{\rm R}}\left(\sin\theta\cos\theta\sin\omega t{\bf x}+\sin^2\theta{\bf y}\right)\\
&&+\frac{e\omega}{\lambda_R}\frac{2\Delta}{\pi\Gamma}\sin\theta\cos\omega t{\bf x},\nonumber
\end{eqnarray}
where $\lambda_{\rm R}=\hbar/(\alpha_{\rm R}m)$ is the Rashba precession length. Equation \eqref{pumpedRash} shows that a dc current is injected transverse to the precession axis and an ac current is generated along it, which is similar to the standard theory, Eq. \eqref{eq:spt1}. The Berry curvature induces both ac and dc responses [first two terms in Eq. \eqref{pumpedRash}] while the Fermi surface breathing produces an ac current along ${\bf x}$, but not along ${\bf y}$ [third term in Eq. \eqref{pumpedRash}]. We emphasize that this expression is correct in the limit of small spin-orbit coupling. Higher harmonics appear at higher orders in spin-orbit coupling.

\subsection{Charge pumping in the magnetic Rashba gas} 

Let us now evaluate the pumped charge current pumped in the magnetic Rashba gas regularized on a hexagonal lattice. It is defined by the Hamiltonian \cite{Ovalle2023}
\begin{eqnarray}
{\cal H}_0=\varepsilon_{\bf k}+\Delta\hat{\bm\sigma}\cdot{\bf m}+\frac{t_{\rm R}}{a}{\bm \eta}_{\bf k}\cdot(\hat{\bm\sigma}\times{\bf z})\label{HR}.
\end{eqnarray}
Explicitly, $\varepsilon_{\bf k}=-2t(\cos{\bf k}\cdot{\bf a}+\cos{\bf k}\cdot{\bf b}+\cos{\bf k}\cdot{\bf c})$, ${\bm \eta}_{\bf k}=2({\bf a}\sin{\bf k}\cdot{\bf a}+{\bf b}\sin{\bf k}\cdot{\bf b}+{\bf c}\sin{\bf k}\cdot{\bf c})$, where ${\bf a}$, ${\bf b}$ and ${\bf c}$ are the lattice vectors connecting the nearest neighbors. Here, $t$ is the nearest-neighbor hopping parameter (fixed to 0.2 in the present work), $\Delta$ is the s-d exchange between the conduction electrons and the localized ones, $t_R$ is the linear Rashba spin-orbit coupling coming from inversion symmetry breaking normal to the ({\bf a}, {\bf b}) plane. Because of the coexistence of s-d exchange and spin-orbit coupling, the band structure and the Fermi surface of the gas are highly sensitive to the magnetization direction. Fig. \ref{fig:sketch}(b) and (c) show the band structure and Fermi surface when the magnetization lies along $+{\bf y}$ (red) and $-{\bf y}$ (black) for two different situations, $\Delta=0.1$ and $t_R=0.1$ (top) and  $\Delta=1$ and $t_R=0.3$ (bottom). Whereas the Fermi surface remains circular in the former case, it gets strongly distorted in the latter case, pointing to different pumping regimes. Short movies of the breathing can be found in Ref. \cite{SuppMat}.

\begin{figure}[h!]
  \includegraphics[width=8cm]{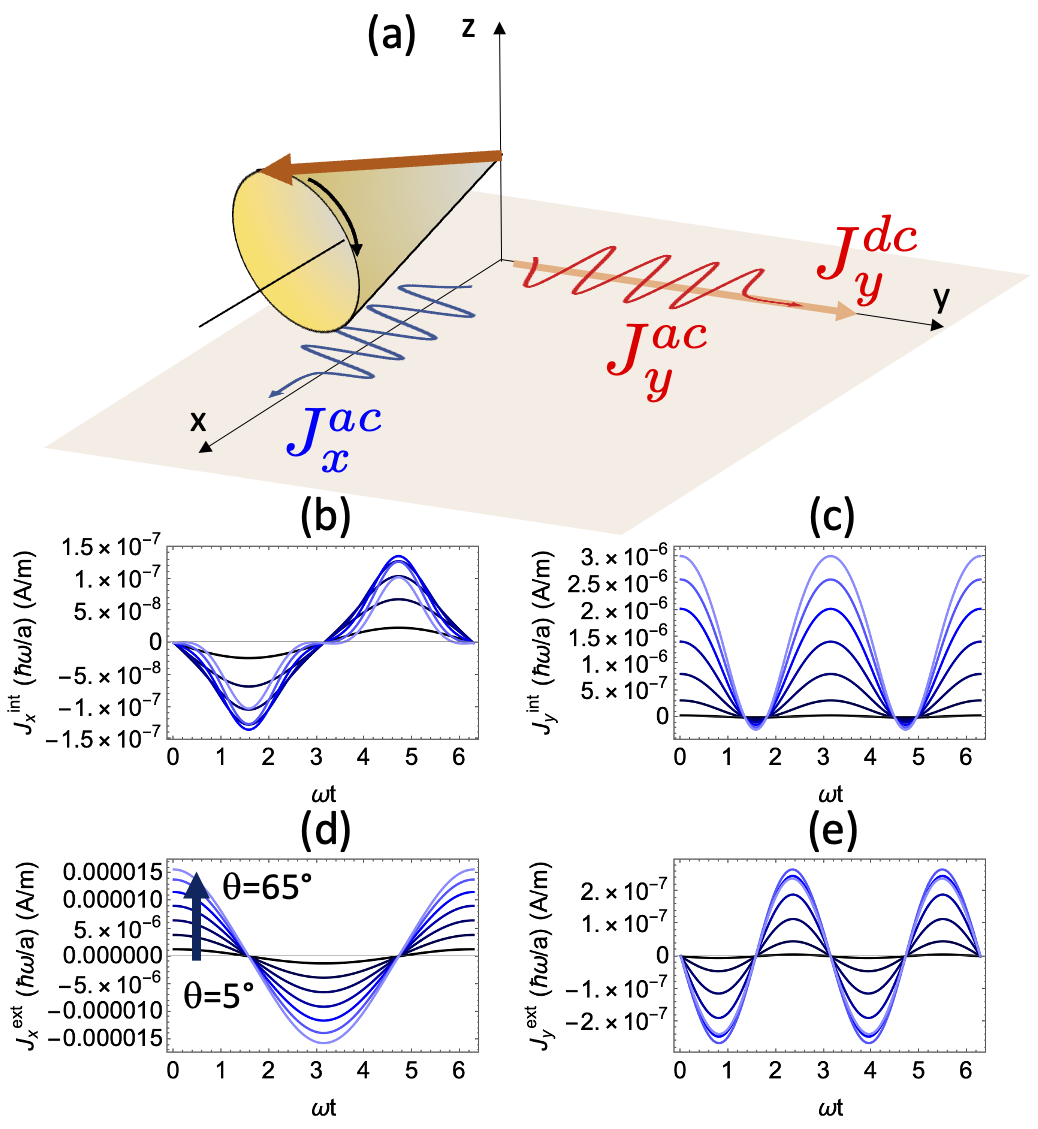}
  \caption{(Color online) (a) Sketch of the charge pumping configuration. Time-dependence of the (b,c) intrinsic currents, $J^{\rm in}_{x,y}$, and (d,e) extrinsic currents, $J^{\rm ext}_{x,y}$, as a function of the cone angle, $\theta\in[5^\circ,65^\circ]$. Both $J_x$ (d,e) and $J_y$ (b,c) are displayed and the parameters are $\Delta=0.1$, $t_R=0.1$, $\Gamma=0.1$ and $E_F=-1$.\label{fig:pump1}}
\end{figure}

We now compute the pumped charge current using Eqs. \eqref{eq:jcsurf} and \eqref{eq:jcsea}. With the definition of ${\bf m}$ given in the previous section, the torque operator is ${\bf T}\cdot\partial_t{\bf m}=\omega\Delta\sin\theta \alpha(t)$, with $\alpha(t)=\sigma_y\cos\omega t+\sigma_z\sin\omega t$. In the following, the energy integral in Eq. \eqref{eq:jcsea} is performed analytically. We start our investigation by adopting the set of parameters $\Delta=0.1$, $t_R=0.1$, for which the Fermi surface remains mostly circular (see Fig. \ref{fig:sketch}). The pumped current components computed for the Rashba gas are displayed in Fig. \ref{fig:pump1}(b-e) and several remarks are in order. First, the signal of all four current components increases steadily with the cone angle, which is expected. Second, whereas Eq. \eqref{eq:spt1} predicts an oscillatory current only along ${\bf x}$, our calculations predict that in the case of the Rashba gas, an oscillatory current also develops along ${\bf y}$, with both extrinsic and intrinsic contributions [Fig. \ref{fig:pump1}(e)]. This distinct feature can be traced back to the competition between the exchange and the Rashba field, depicted in Fig. \ref{fig:sketch}, and which results in a time-dependent modulation of the effective field ${\bf B}_{\rm eff}$ along ${\bf x}$ that produces the oscillatory current along ${\bf y}$. By symmetry, the current along ${\bf x}$ possesses odd harmonics, $(2n+1)\omega$, whereas the current along ${\bf y}$ possesses even harmonics, $2n\omega$. Only $J_y$ produces a rectified, dc current. 
 In addition, the intrinsic current along {\bf x} ({\bf y}) displays a $\sin\omega t$ ($\cos 2\omega t$) dependence whereas the extrinsic current along {\bf x} (${\bf y}$) displays a $\cos\omega t$ ($\sin 2\omega t$) dependence. This phase shift between intrinsic and extrinsic currents is also present in the conventional theory, Eq. \eqref{eq:spt1}, where $J_{c,x}\propto{\tilde G}^r_{\uparrow\downarrow}\sin\omega t+{\tilde G}^i_{\uparrow\downarrow}\cos\omega t$. Finally, we find that the extrinsic contribution to $J_x$ is much larger than the intrinsic one, typically two orders of magnitudes in Fig. \ref{fig:pump1}(b,d). The extrinsic contribution is, by definition, inversely proportional to the disorder broadening $\Gamma$, which in our model is a tunable parameter, fixed to $\Gamma=0.1$ eV. Decreasing this parameter would lead to an enhancement of the extrinsic contribution. The intrinsic contribution, on the other hand, is related to the time-momentum Berry curvature, and is therefore expected to be very sensitive to avoided crossing points in the band structure, akin to anomalous Hall effect (see, for instance, discussions in Ref. \cite{Nagaosa2010}). Therefore, the relative magnitude of the intrinsic and extrinsic contributions is not only band structure-dependent but also disorder-dependent, which opens particularly appealing avenues for charge pumping engineering in quantum materials, such as magnetic Weyl semimetals.\par

\begin{figure}[h!]
  \includegraphics[width=8cm]{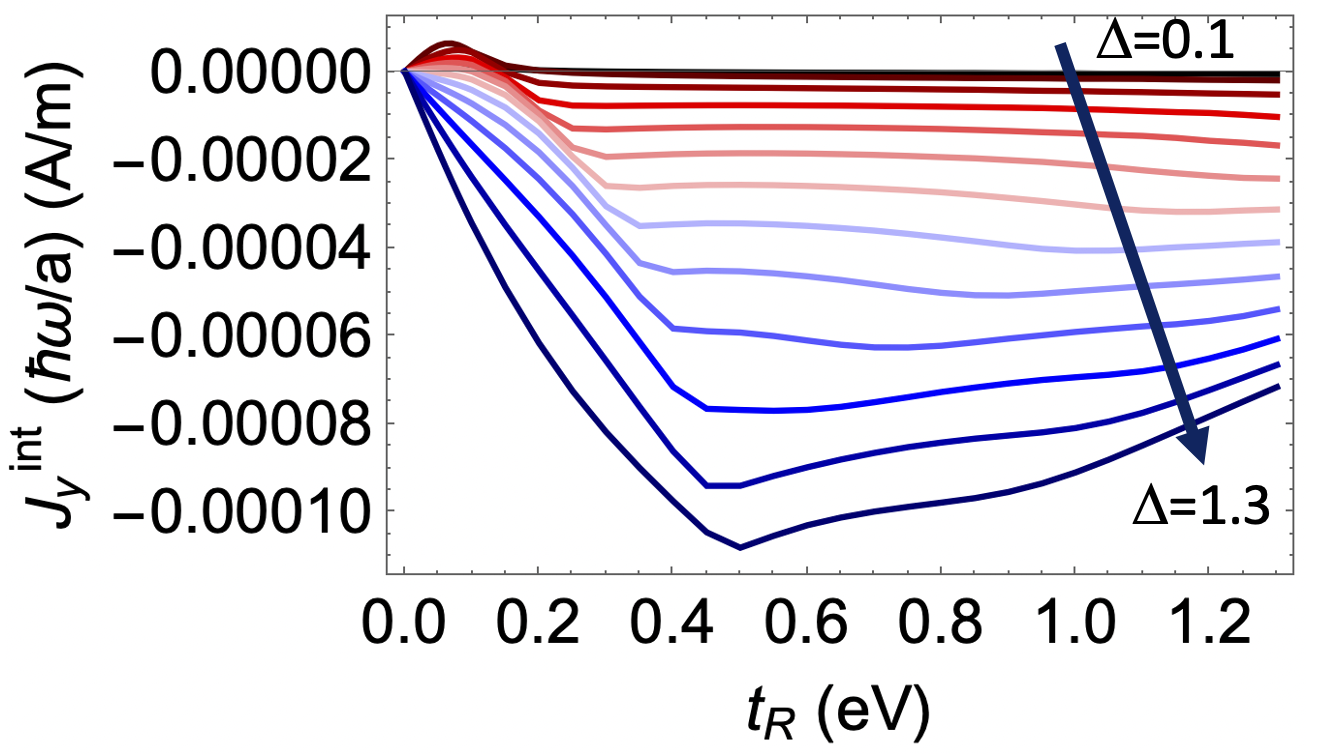}
  \caption{(Color online) Dependence of the (a) intrinsic and (b) extrinsic dc current along ${\bf y}$ as a function of the Rashba and exchange interactions. The parameters are $\Delta\in[0.1,1.3]$, $t_R\in[0,1.3]$, $\Gamma=0.1$, $E_F=-1$ and $\theta=65^\circ$. \label{fig:pump3}}
\end{figure}

\begin{figure}[h!]
  \includegraphics[width=8.5cm]{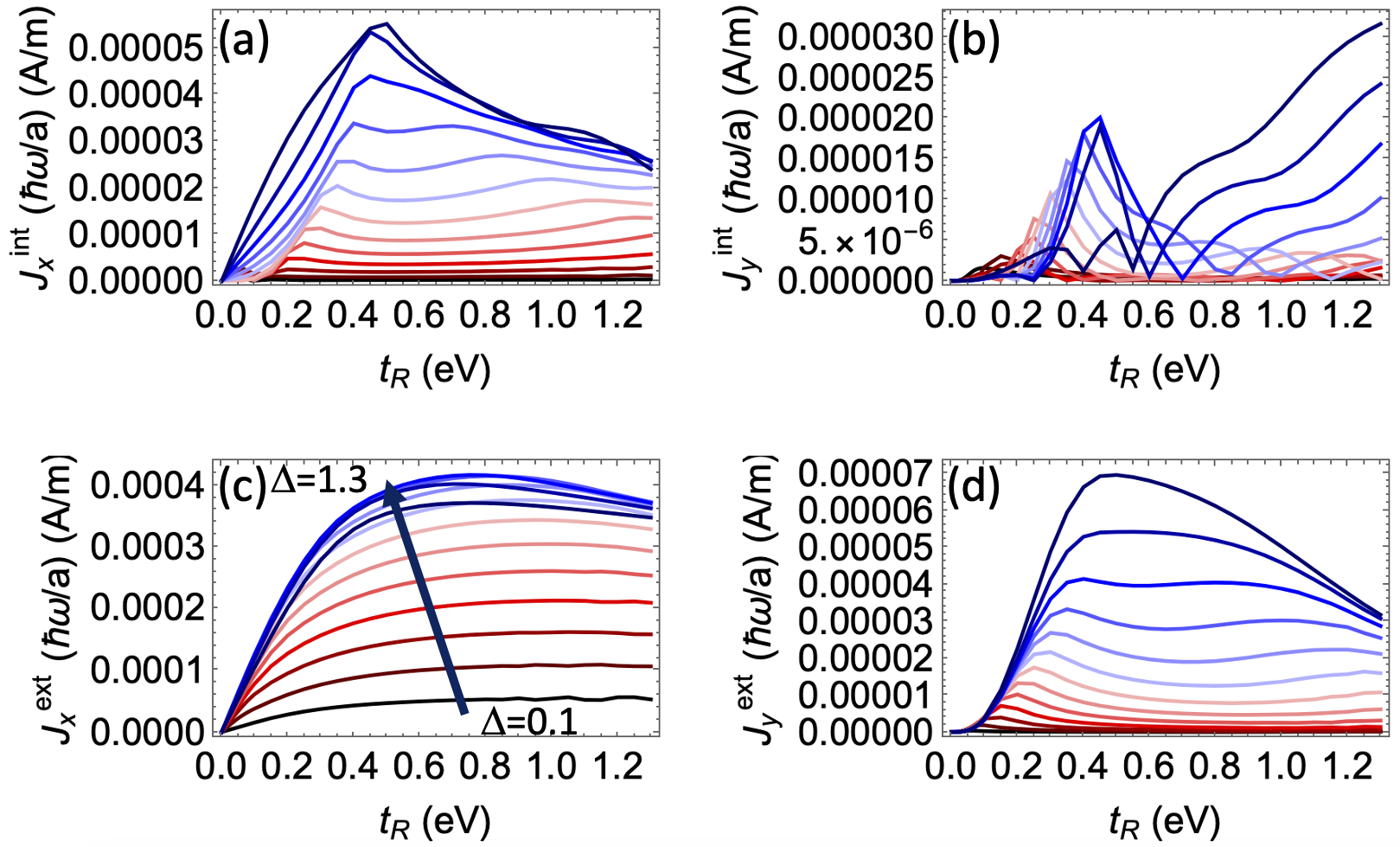}
  \caption{(Color online) Dependence of the ac current along (a,c) ${\bf x}$ and (b,d) ${\bf y}$ as a function of the Rashba and exchange interactions. Both (a,b) intrinsic and (c,d) extrinsic contributions are shown. The parameters are the same as in Fig. \ref{fig:pump3}. \label{fig:pump4}}
\end{figure}

We now investigate the influence of the relative strength of the Rashba and exchange interactions on the dc (Fig. \ref{fig:pump3}) and ac (Fig. \ref{fig:pump4}) currents. The dc current, $J_y^{\rm int}$, displays two interesting features. First, the overall magnitude of the pumped current increases with the exchange $\Delta$, which is expected from the linear response theory. Second, and more interestingly, the pumped dc current depends on the Rashba strength $t_R$ in a nontrivial manner. At small Rashba strength ($t_R\rightarrow0$), the current is proportional to $t_R$, and its slope is either positive ($\Delta\leq0.6$) or negative ($\Delta\geq0.6$) depending on the exchange parameter. Since $J_y^{\rm int}$ is associated with interband transitions via the Berry curvature, it is highly sensitive to the relative positions of the bands. For $\Delta\geq0.6$, the two bands are well separated, leading to a negative dc current for all values of $t_R$. For intermediate Rashba strength, the current increases steadily up to a maximum. Once the maximum is reached, the magnitude of the pumped dc current saturates and decreases smoothly. This behavior is obviously in stark contrast with the conventional theory of spin pumping and illustrates the complex interplay between exchange, Rashba field, and spin-to-charge conversion.

To complete the physical picture, Fig. \ref{fig:pump4} displays the magnitude of the lowest harmonic of the ac charge current as a function of the strength of exchange and Rashba interactions. $J_x\sim \cos(\omega t+\phi_0)$ is reported in Figs. \ref{fig:pump4}(a,c) and $J_y\sim \cos(2\omega t+\phi_0)$ is given in Figs. \ref{fig:pump4}(b,d). We emphasize that we show the {\em absolute value} of the ac components. The extrinsic contributions to $J_x$ and $J_y$ [Fig. \ref{fig:pump4}(c,d)] and the intrinsic contribution to $J_x$ [Fig. \ref{fig:pump4}(a)] increases steadily with the Rashba strength, saturates, and smoothly decreases at large spin-orbit interaction, similarly to the dc case. In contrast, the intrinsic contribution to $J_y$ [Fig. \ref{fig:pump4}(b)] reaches a maximum, collapses, and increases again at large Rashba strength. Since we only account for the absolute value of the ac current, the collapse observed in Fig. \ref{fig:pump4}(b) is associated with a $\pi$ shift (i.e., a sign change).

\begin{figure}[h!]
  \includegraphics[width=9cm]{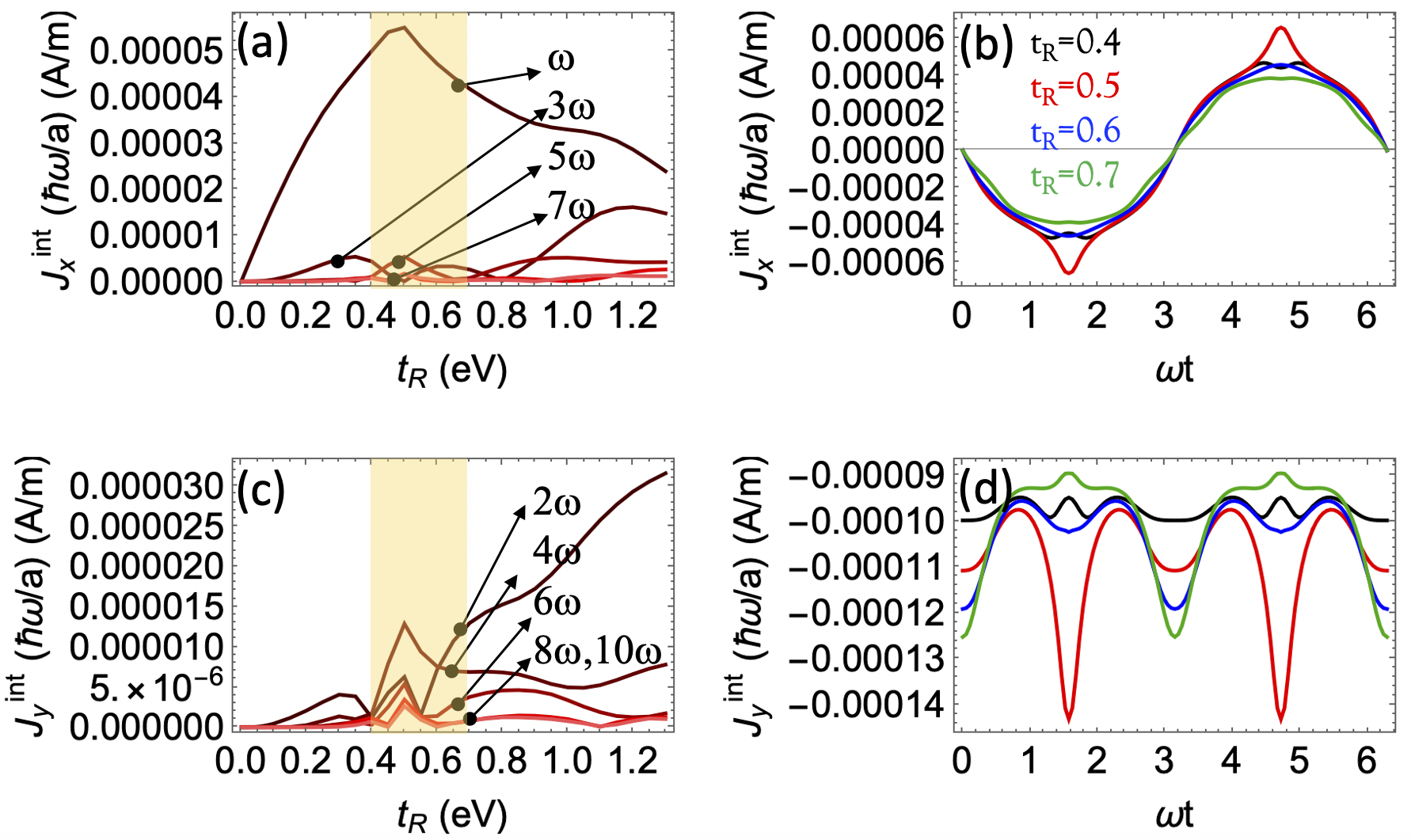}
  \caption{(Color online) Dependence of the current harmonics as a function of the Rashba strength for $\Delta=1$ for the intrinsic contribution of the current along (a) ${\bf x}$ and (c) ${\bf y}$. The shaded area emphasizes the parameter region when harmonics are sizable. Time-dependence of the intrinsic current along (b) ${\bf x}$ and (d) ${\bf y}$ for $t_R=0.2,\;0.3,\;0.4$ and 0.5. \label{fig:pump5}}
\end{figure}

As discussed above, the competition between exchange and spin-orbit coupling leads to a regime of parameters where the Berry curvature in time-momentum space induces higher harmonics. Importantly, we find that the harmonics are particularly strong for the intrinsic current contributions and rather negligible in the extrinsic contribution. In the following, we focus on the intrinsic currents. Figure \ref{fig:pump5} shows such a situation, obtained when $\Delta=1$, $E_F=-1$, and $\theta=65^\circ$. Although their magnitude decreases with the harmonics number (under the adiabatic assumption, our theory is based on a perturbative expansion of the magnetization dynamics), we find that in a certain region of Rashba strength, close to the maximum dc current obtained in Fig. \ref{fig:pump3}(a), i.e., $t_R\in[0.3,0.6]$, the first few harmonics are comparable in magnitude. This results in a rather complex time-dependence of the intrinsic signal, as shown in Figure \ref{fig:pump5}(b,d) for selected cases. 

\subsection{Magnetic graphene} 

\begin{figure}[h!]
  \includegraphics[width=9cm]{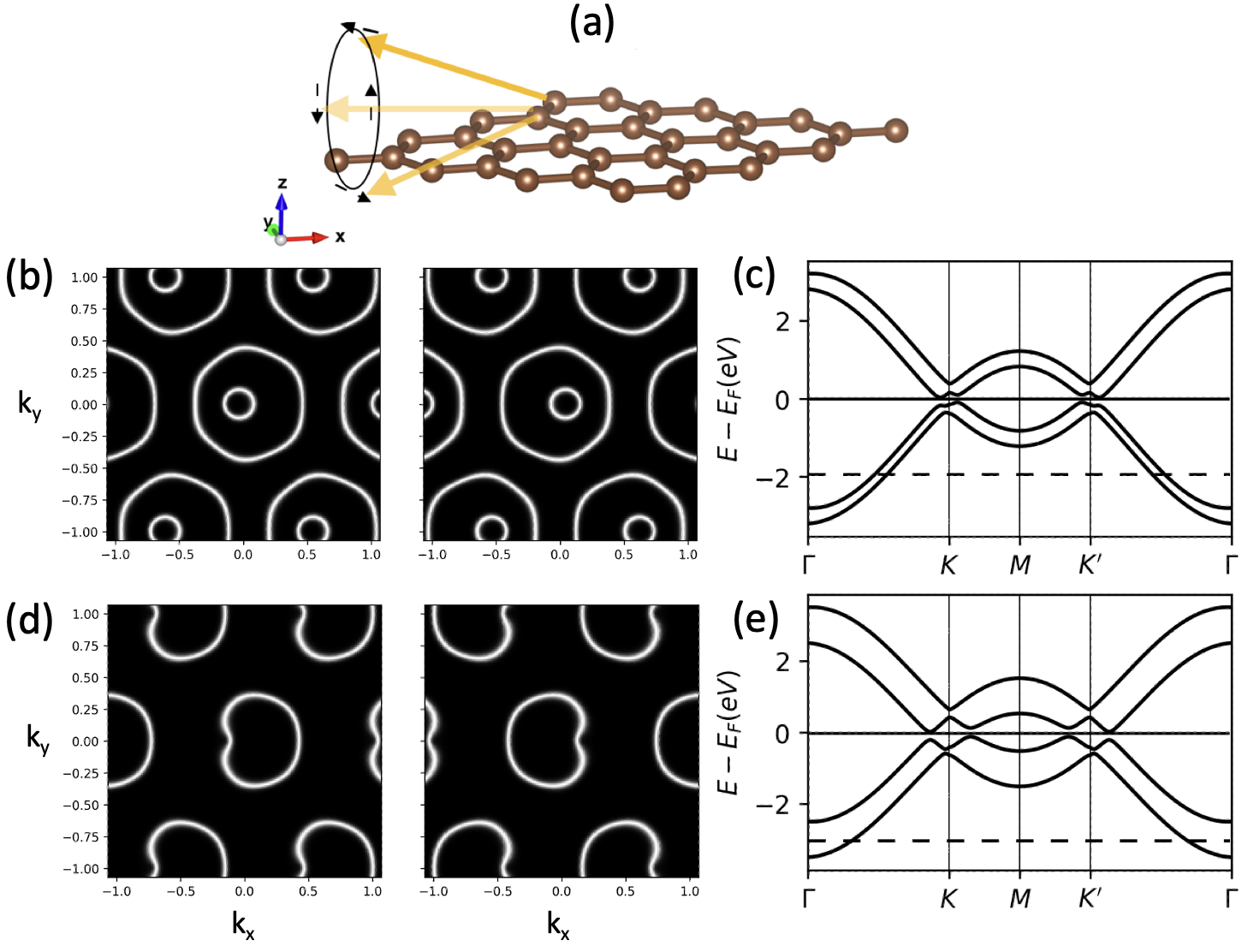}
  \caption{(Color online) (a) Schematics of spin pumping in magnetic graphene with Rashba spin-orbit coupling. (b,d) Fermi surface for $\Delta=0.1$ eV ($\Delta=0.75$ eV) and $t_R=0.15$ eV when setting the magnetization along $\pm{\bf x}$. (c,e) Corresponding band structure when setting the magnetization along ${\bf z}$. The dashed lines correspond to the energy at which the surfaces in (b,d) are taken.\label{fig:pump6}}
\end{figure}

We now consider the case of magnetic graphene, Fig. \ref{fig:pump6}(a), whose band structure presents the peculiarity to be highly sensitive to the magnetization direction. The Hamiltonian is obtained by regularizing Hamiltonian \eqref{HR} on a honeycomb lattice. Setting $\theta=45^\circ$, $t=1$ eV, and $t_R=0.15$ eV, we obtain the band structures reported in Fig. \ref{fig:pump6}(c,e) for $\Delta=0.1$ eV and $\Delta=0.75$ eV, respectively. For a better comparison with the Rashba gas studied above, we set the lattice parameter to $a=1$ nm. Figures \ref{fig:pump6}(b,d) show the Fermi surface when setting the magnetization along $\pm{\bf x}$ for two different illustrative cases. Because the breathing is governed by the Rashba spin-orbit coupling, we obtain a distortion that is qualitatively similar to the one in Fig. \ref{fig:sketch}, suggesting strong charge pumping. 

\begin{figure}[h!]
  \includegraphics[width=9cm]{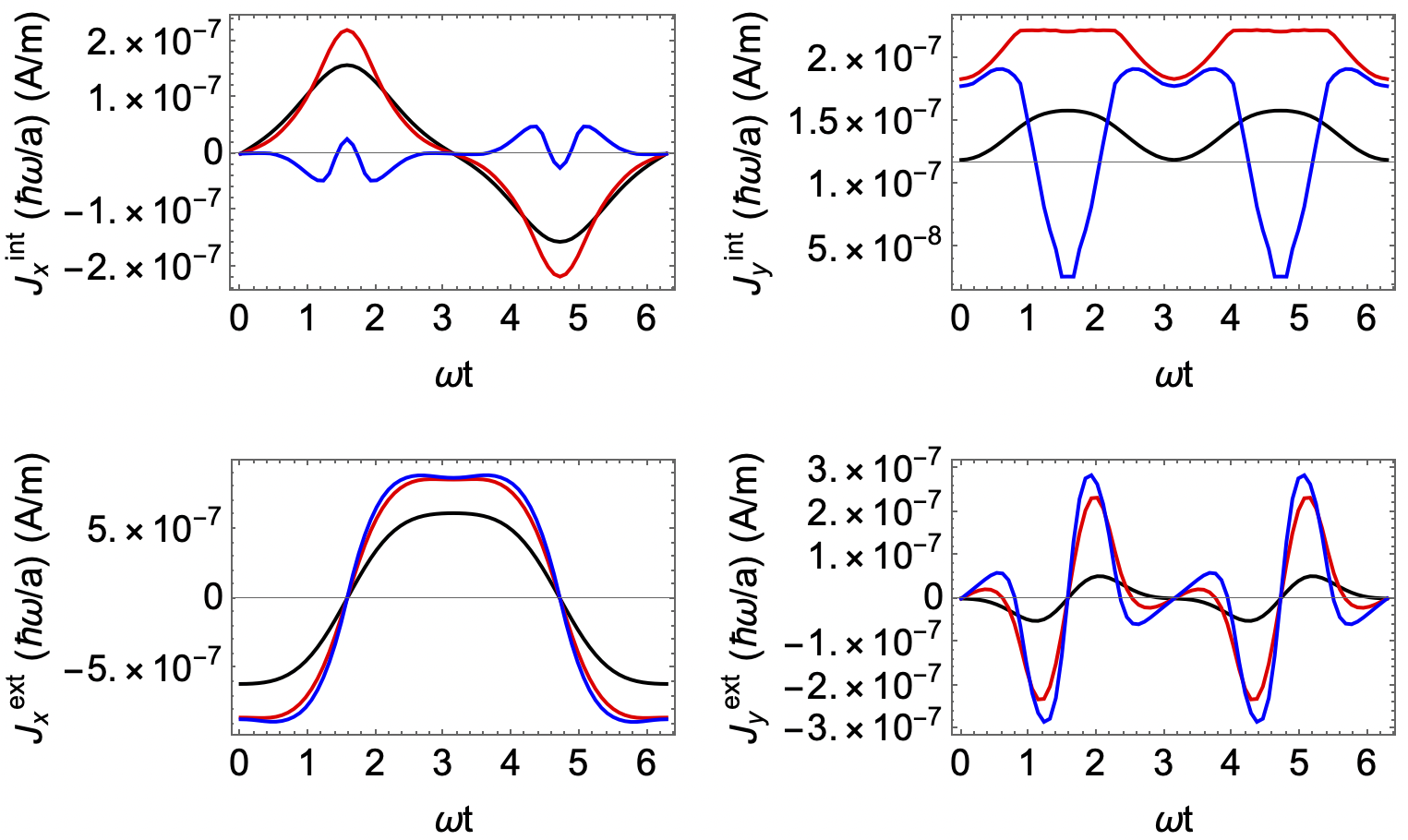}
  \caption{(Color online) Time-dependence of the intrinsic (a,b) and extrinsic (c,d) currents for $t=0.2$ eV, $\Delta=1.0$ eV, $E_F=-0.5$ eV and $t_R=0.3$ (black), 0.5 (red) and 0.7 eV (blue). \label{fig:pump7}}
\end{figure}

The time-dependence of the intrinsic and extrinsic currents are reported in Fig. \ref{fig:pump7} for parameters comparable to that of the Rashba gas, i.e., $t=0.2$ eV, $\Delta=1.0$ eV, $E_F=-0.5$ eV and $t_R=0.3$ (black), 0.5 (red) and 0.7 eV (blue). For this set of parameters, the time-dependence of the pumped current is radically different from the conventional one reported in Fig. \ref{fig:pump1}, and substantially deviates from a harmonic response (i.e., $\cos \omega t$, $\sin \omega t$). Remarkably, these higher harmonics appear in both extrinsic and intrinsic current contributions. This behavior is directly associated with the high sensitivity of the band structure on the magnetization, as seen in Fig. \ref{fig:pump6}, and results in high harmonics.

\begin{figure}[h!]
  \includegraphics[width=9cm]{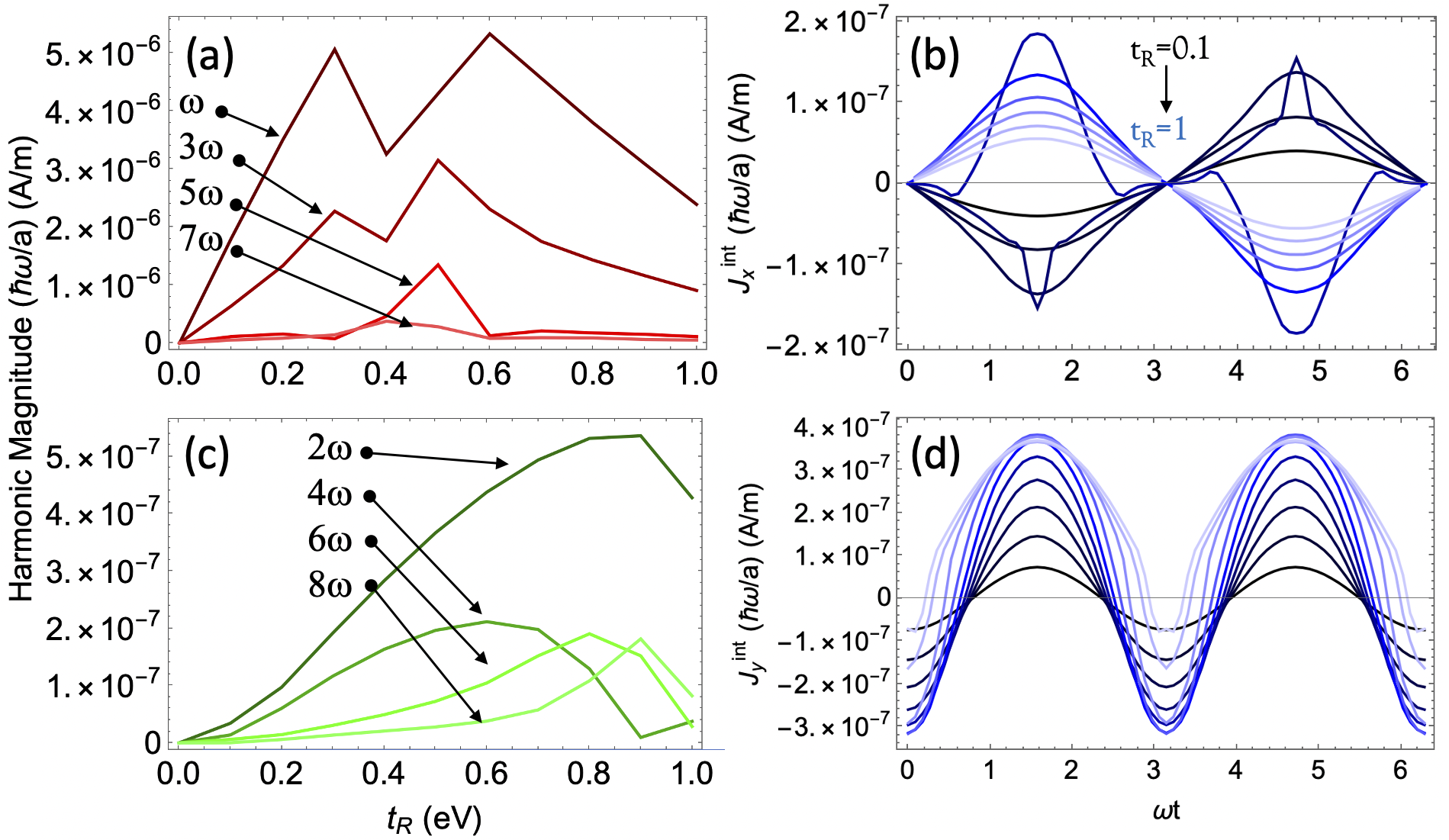}
  \caption{(Color online) Dependence of the current harmonics as a function of the Rashba strength for $t=1$ eV, $\Delta=0.75$ eV and $E_F=-1.15$ eV for the intrinsic contribution of the current along (a) ${\bf x}$ and (c) ${\bf y}$. Time-dependence of the intrinsic current along (b) ${\bf x}$ and (d) ${\bf y}$ for $t_R\in[0.1,1]$ eV. \label{fig:pump8}}
\end{figure}

Finally, taking a hopping parameter closer to that of graphene, $t=1$ eV, Fig. \ref{fig:pump8} shows the dependence of the harmonics of the intrinsic current, obtained when $\Delta=0.75$ eV, $E_F=-1.15$ eV, and $\theta=45^\circ$. As expected, the current along ${\bf x}$ possesses odd harmonics ($(2n+1)\omega$) whereas the current along ${\bf y}$ possesses even harmonics ($2n\omega$). Upon increasing the Rashba strength, the homodyne component along  ${\bf x}$ increases steadily up to a maximum that lies out of the range studied here. The other harmonic components all reach a maximum within the range $t_R\in[0,1]$ eV suggesting that second and third harmonic are quite sizable for a reasonable Rashba strength, resulting in the complex time-dependence of the intrinsic signal, see Fig. \ref{fig:pump8}(b,d).

\section{Discussion and Conclusion} 

The formalism described in the present Article extends the traditional theory of spin and charge pumping \cite{Brataas2002,Tserkovnyak2002b} by covering the full range of exchange and spin-orbit coupling. It can be readily adapted to address spin, charge, and orbital pumping in multiband heterostructures. It is particularly well adapted to compute adiabatic pumping in realistic heterostructures obtained from density functional theory. Among the important features uncovered by the present theory, we point out the importance of both intrinsic and extrinsic contributions to the ac currents, akin to magnetic damping \cite{Gilmore2007}, an aspect that is overlooked by the traditional theory of spin pumping. This theoretical framework is instrumental to investigate spin-charge interconversion in strongly spin-orbit coupled systems such as the surface of topological heterostructures, involving topological insulators and Weyl semimetals, for instance. 

\begin{figure}[h!]
  \includegraphics[width=8cm]{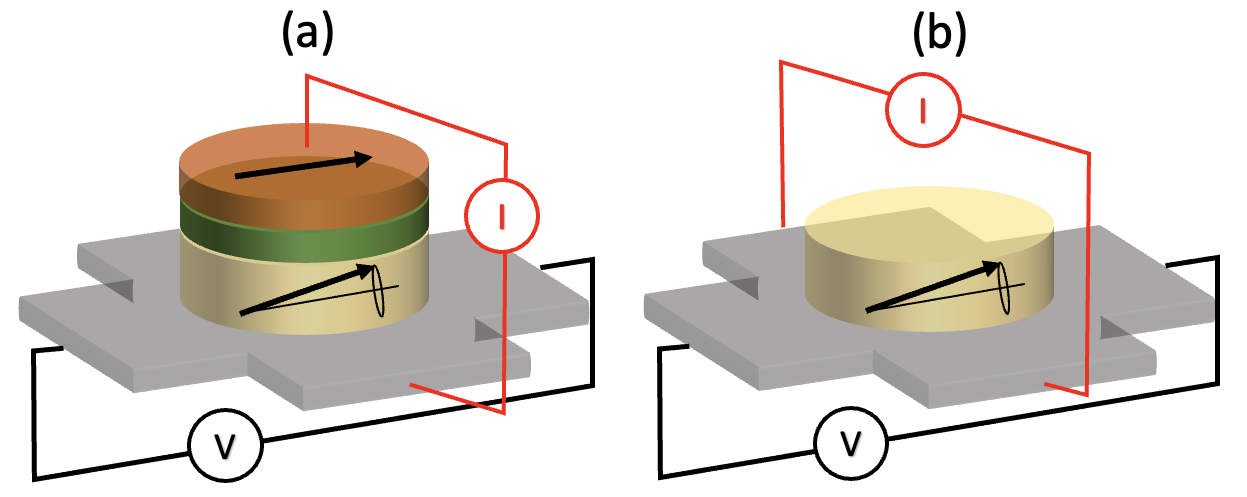}
  \caption{(Color online) Two devices implementing charge pumping induced by large angular precession. In device (a), the magnetization autooscillation is induced by spin transfer torque, whereas in device (b) it is induced by spin-orbit torque. Notice that in both cases, the pumped current is collected {\em along} the magnetization direction (black connectors), decoupled from the driving current (red connectors). \label{fig:pump9}}
\end{figure}

Finally, we would like to comment on the adiabatic harmonic generation displayed by Figs. \ref{fig:pump5} and \ref{fig:pump8}. In contrast with Ref. \onlinecite{Ly2022} that displays tens of harmonics, the adiabatic theory only shows a few of them. The present theory differs from Ref. \onlinecite{Ly2022} in several aspects: first Ref. \onlinecite{Ly2022} computes the time-dependent Schr\"odinger equation in a finite size ribbon in the Landauer-B\"uttiker configuration (a conductor connected to two leads). As such, the real-time dynamics of the electron spin is computed numerically and the quantum interference between the electronic modes contributes to the complex time-dependent current. In the present theory, it is assumed that the spin is aligned on the effective field (exchange field+Rashba field, ${\bf B}_{\rm eff}$ in Fig. \ref{fig:sketch}) so that the internal spin dynamics is neglected. In addition, our model Hamiltonian is not defined in real space but in momentum space. It is therefore unsurprising that several features associated with quantum interference and real-time spin evolution are absent in the present work.

Our calculations suggest that in the case of a Rashba gas, these harmonics are mostly present in the intrinsic contribution, associated with the time-momentum Berry curvature, whereas in the case of magnetic graphene, it is present in both intrinsic and extrinsic contributions. This distinct behavior suggests that energy dispersion plays a crucial role. A natural research direction is therefore to identify materials systems where the Berry curvature is enhanced, e.g. in magnetic Weyl semimetals such as Co$_3$Sn$_2$S$_2$ \cite{Liu2019d,Morali2019} or Mn$_3$Sn \cite{Kuroda2017}. In addition, the harmonics computed here are obtained for {\em large precession angles}, which are not achievable using ferromagnetic resonance techniques. Nonetheless, such large angles can be obtained via current-driven autooscillations \cite{Kiselev2003,Liu2012b}. This inspires us to propose two devices in Fig. \ref{fig:pump9} based on (a) spin transfer torque and (b) spin-orbit torque that can excite such large-angle auto-oscillations. In the presence of strong spin-orbit coupling, large Fermi surface breathing is expected, which could trigger harmonic currents. 

We acknowledge that the generation of harmonic currents via spin pumping is a daring technical challenge. In fact, to the best of our knowledge, only a handful of experiments have achieved the detection of the homodyne ac current \cite{Hahn2013,Weiler2014,Wei2014}. The difficulty is that not only does one need to get rid of the frequency response of the radiofrequency setup itself, but in addition, in most experiments the magnetization precession is not circular but rather ellipsoidal, resulting in harmonics of purely magnetic origin \cite{Hahn2013}. Hence, one needs to ensure that the observed harmonics are of purely electronic nature, i.e., from the competition of exchange and spin-orbit coupling, which is certainly an interesting challenge for experiments.

\begin{acknowledgments}
A.P. acknowledges support from the ANR ORION project, grant ANR-20-CE30-0022-01 of the French Agence Nationale de la Recherche. A. M. acknowledges support from the Excellence Initiative of Aix-Marseille Universit\'e - A*Midex, a French "Investissements d'Avenir" program.\end{acknowledgments}
\bibliography{Biblio2023}

\end{document}